\documentclass[preprint]{emulateapj}
\usepackage{amsmath,amssymb}
\usepackage{graphicx}
\usepackage{natbib}

\usepackage{aas_macros} %added later
\slugcomment{}
\shortauthors{Hirano et al.}
\shorttitle{Global Analysis of KOI-977}
\begin{document}
%%%%%%%%%%%%%%%%%%%%%%%%%%%%%%%%%%%%%%%%%%%%%%%%%%%%%%%%%%%%%%%%%%%%%%
%%%%%%%%%%%%%%%%%%%%%%%%%%%%%%%%%%%%%%%%%%%%%%%%%%%%%%%%%%%%%%%%%%%%%%
\title{Global Analysis of KOI-977: Spectroscopy, Asteroseismology, and Phase-curve Analysis}
%%%%%%%%%%%%%%%%%%%%%%%%%%%%%%%%%%%%%%%%%%%%%%%%%%%%%%%%%%%%%%%%%%%%%%
%%%%%%%%%%%%%%%%%%%%%%%%%%%%%%%%%%%%%%%%%%%%%%%%%%%%%%%%%%%%%%%%%%%%%%
\author{
Teruyuki Hirano\altaffilmark{1}, 
Kento Masuda\altaffilmark{2}, 
Bun'ei Sato\altaffilmark{1}, 
Othman Benomar\altaffilmark{3},
Yoichi Takeda\altaffilmark{4},
Masashi Omiya\altaffilmark{4},
Hiroki Harakawa\altaffilmark{4},
and Atsushi Kobayashi\altaffilmark{1}
} 
%%%%%%%%%%%%%%%%%%%%%%%%%%%%%%%%%%%%%%%%%%%%%%%%%%%%%%%%%%%%%%%%%%%%%%
\altaffiltext{1}{Department of Earth and Planetary Sciences, Tokyo Institute of Technology,
2-12-1 Ookayama, Meguro-ku, Tokyo 152-8551, Japan}
\altaffiltext{2}{Department of Physics, The University of Tokyo, Tokyo 113-0033, Japan}
\altaffiltext{3}{Department of Astronomy, The University of Tokyo, Tokyo 113-0033, Japan}
\altaffiltext{4}{National Astronomical Observatory of Japan, 
2-21-1 Osawa, Mitaka, Tokyo, 181-8588, Japan}
%\altaffiltext{4}{Department of Physics, and Kavli Institute for Astrophysics and 
%Space Research, Massachusetts Institute of Technology, Cambridge, MA 02139}
%\altaffiltext{4}{Department of Astronomy, The University of Tokyo, 
%Tokyo, 113-0033, Japan}
%\altaffiltext{5}{Research Center for the Early Universe, School of Science, 
%The University of Tokyo, Tokyo 113-0033, Japan}
%\altaffiltext{6}{Department of Astrophysical Sciences, 
%Princeton University, Princeton, NJ 08544}

\email{hirano@geo.titech.ac.jp}
%%%%%%%%%%%%%%%%%%%%%%%%%%%%%%%%%%%%%%%%%%%%%%%%%%%%%%%%%%%%%%%%%%%%%%
\begin{abstract}
We present a global analysis of KOI-977, one of the planet host candidates detected by
{\it Kepler}. Kepler Input Catalog (KIC) reports that KOI-977 is a red giant, for 
which few close-in planets have been discovered. 
Our global analysis involves spectroscopic and asteroseismic determinations
of stellar parameters (e.g., mass and radius) and radial velocity (RV) measurements. 
Our analyses reveal that KOI-977 is indeed a red giant in the red clump, but its estimated 
radius ($\gtrsim 20R_\odot=0.093$ AU) is much larger than KOI-977.01's orbital distance
($\sim 0.027$ AU) estimated from its period ($P_\mathrm{orb}\sim 1.35$ days) and host star's mass. 
RV measurements show a small variation, which also contradicts
the amplitude of ellipsoidal variations seen in the light-curve folded
with KOI-977.01's period. 
Therefore, we conclude that KOI-977.01 is a false positive, meaning that the red giant, 
for which we measured the radius and RVs, is different from the object 
that produces the transit-like signal (i.e., an eclipsing binary). 
On the basis of this assumption, we also perform a light-curve analysis including the modeling 
of transits/eclipses and phase-curve variations, adopting various values for the dilution factor
$D$, which is defined as the flux ratio between the red giant and eclipsing binary. 
Fitting the whole folded light-curve as well as individual transits in the short cadence data simultaneously, 
we find that the estimated mass and radius ratios of the eclipsing binary are consistent with 
those of a solar-type star and a late-type star (e.g., an M dwarf) for $D\gtrsim 20$. 
\end{abstract}
%%%%%%%%%%%%%%%%%%%%%%%%%%%%%%%%%%%%%%%%%%%%%%%%%%%%%%%%%%%%%%%%%%%%%%
%%%%%%%%%%%%%%%%%%%%%%%%%%%%%%%%%%%%%%%%%%%%%%%%%%%%%%%%%%%%%%%%%%%%%%

\section{Introduction\label{s:intro}}\label{s:intro}
Transiting planets have attracted much recent attention thanks
to their ability to characterize planetary orbits, atmospheres, 
and evolution history. 
In particular, the {\it Kepler} spacecraft \citep{2010Sci...327..977B} has yielded a wide variety 
of planetary systems including notably unique ones
\citep[e.g.,][]{2011Sci...333.1602D, 2011Natur.470...53L, 2012ApJ...752....1R, 
2012ApJ...759L..36H, 2013Sci...342..331H, 2014Sci...344..277Q}. 
%such as circumbinary planets around eclipsing binaries 
%\citep[e.g., Kepler-15;][]{2011Sci...333.1602D}, 
%an ultra-short period planet releasing its atmosphere and dust 
%\citep[e.g., KIC 12557548;][]{2012ApJ...752....1R}. 

One nuisance concerning searches for transiting planets is that 
transit photometry is often contaminated by background/foreground
sources, and such contaminators are sometimes responsible for 
the periodic light-curve depressions that mimic planetary transits 
(so-called false positives). According to the recent statistical study, the false positive 
rate for Kepler Objects of Interest (KOI) is estimated to be $7-18\%$ depending on 
the size of planet candidates \citep{2013ApJ...766...81F}. 
The false positive rate becomes
vanishingly small when we focus only on multiple transiting 
systems \citep{2012ApJ...750..112L}, but it is still necessary for their confirmations 
to use independent techniques such as radial velocity (RV) measurements or 
transit timing variations (TTVs).

In this paper, we present a global analysis for KOI-977, one of the
planet-host candidates reported by {\it Kepler} \citep{2013ApJS..204...24B}
with a goal of confirming the planetary nature of KOI-977.01. 
Kepler Input Catalog (KIC) reports that KOI-977 (KIC 11192141)
is an evolved star (giant) with the effective temperature of $T_\mathrm{eff}=4283$ K 
and surface gravity of $\log g=2.012$. 
Later studies confirmed KOI-977 to be a red giant from a medium-resolution 
spectroscopy in the near infrared region \citep{2012ApJ...750L..37M, 2014ApJS..213....5M}
or photometric analysis \citep{2013ApJ...767...95D}. 
Contrary to the situation for main sequence stars, 
only a few hot Jupiters are so far reported around evolved stars 
\citep[e.g.,][]{2010ApJ...721L.153J, 2013MNRAS.428.1077S, 2014A&A...562A.109L}
in spite of many dedicated RV surveys for those stars. 
The lack of close-in planets is often attributed to the outcome 
of tidal interactions \citep[e.g.,][]{2008PASJ...60..539S, 2013ApJ...772..143S, 2014A&A...566A.113J}, 
but there has been a debate
on this paucity \citep[e.g.,][]{2011ApJ...737...66K}. In this context, a detection or non-detection 
of a hot Jupiter around KOI-977 should become an important constraint on 
the occurrence rate and/or tidal evolution of giant planets around evolved stars.

With a quick look at the light-curve for KOI-977, we noticed that  
the light-curve folded with KOI-977.01's period (Figure \ref{fig:phasebest}) shows a clear pattern 
of well-known flux modulations by ellipsoidal variation and emission/reflected
light from KOI-977.01 \citep[e.g.,][]{2011MNRAS.415.3921F}. Indeed, giants/subgiants 
are ideal targets for phase-curve analyses since the amplitude of ellipsoidal variation 
is approximately proportional to the cube of stellar radius \citep[e.g.,][]{2011AJ....142..195S}. 
The short orbital period of KOI-977.01 ($P_{\rm orb}=1.3537763 \pm 0.0000063$ days), 
along with the moderate transit depth ($\sim1.37$ mmag), 
makes this system an interesting target for further observations/analyses.

Our global analysis involves spectroscopic and asteroseismic determinations
of host star's properties, RV measurements, and light-curve analysis
of the public {\it Kepler} data for KOI-977. In Sections \ref{s:obs} and \ref{s:seismology}, 
we show via spectroscopy and asteroseismology that the most luminous
object in KOI-977 is indeed a red giant in the red clump, but its estimated radius 
is much larger than the estimated semi-major axis of KOI-977.01, inferred from Kepler's third law. 
The observed RVs taken by High Dispersion Spectrograph (HDS) on
the Subaru 8.2m telescope show a small variation, and 
the best-fit model suggests a very eccentric orbit, which is
incompatible with the phase-folded light-curve for KOI-977.01
%with the expected amplitude estimated from KOI-977.01's radius 
%on the assumption that the giant is orbited by KOI-977.01 
(Section \ref{s:obs}). 
Therefore, we are obliged to conclude that KOI-977.01 is likely a false positive, meaning
that the pulsating giant (for which we took RVs) is different from the object 
yielding the transit-like signal (i.e., eclipsing binary). 
Based on this assumption, we present the light-curve analysis in order to put a 
constraint on the properties of the eclipsing binary in Section \ref{s:model}. Employing the EVIL-MC
model by \citet{2012ApJ...751..112J} with some revisions, we simultaneously model the phase-curve 
variation (i.e., ellipsoidal variation, Doppler boosting, and emission/reflected light from the companion)
and transit/eclipse light-curve. Section \ref{s:discussion} is devoted to discussion and summary.

%%%%%%%%%%%%%%%%%%%%%%%%%%%%%%%%%%%%%%%%%%%%%%%%%%%%%%%%%%%%%%%%%%%%%%
\section{Observations and Spectroscopic Analysis\label{s:obs}}\label{s:obs}
\subsection{Observations and RV Measurements\label{s:rv}}\label{s:rv}
We conducted spectroscopic observations on (UT) 2013 June 29, July 1, and
2014 July 13, 14 with Subaru/HDS and obtained RVs for KOI-977.  
We employed the ``I2b" setup in the 2013 campaign and ``I2a" setup 
in 2014 with Image Slicer \#2, with which we can achieve 
a spectral resolution of $R\sim 80000$ \citep{2012PASJ...64...77T}. 
The spectra for RV measurements were taken with the iodine (I$_2$) 
cell for a precise wavelength calibration. On July 12 (2014), we also took 
high signal-to-noise ratio (SNR) spectra without the I$_2$ cell for an RV 
template as well as to estimate basic spectroscopic parameters.

The raw spectra were reduced by the standard IRAF procedure; they were 
bias-subtracted, flat-fielded, and subjected to the scattered light subtraction
before extracting one-dimensional spectra. 
All spectra were subjected to the wavelength calibration by 
Th-Ar emission spectra taken during the twilights, but for the exposures with I$_2$ cell, 
these rough wavelength solutions were later recalibrated based on I$_2$ absorption 
lines for a more precise calibration for the RV measurement \citep{2002PASJ...54..873S}.
The resulting SNR for KOI-977 was typically $\sim100$ 
per pixel around sodium D lines.

Following the RV analysis pipeline by \citet{2002PASJ...54..873S}, we compute
the RVs for KOI-977. First, we estimate the instrumental profile (IP)
for the night during which the stellar template spectrum was taken,
from the flat-lamp spectrum transmitted through the I$_2$ cell. We then deconvolve
the high SNR template with the IP and extract the intrinsic stellar spectrum. 
Each of the observed spectra with the I$_2$ cell is modeled and fitted 
using this intrinsic spectrum with the relative RV and other relevant parameters describing 
the observed spectrum (e.g., those for IP and relative depth of the I$_2$ absorption lines)
being free. The result is summarized in Table \ref{tab:rv}; the RV precision (statistical error) 
is typically $\sim 4$ m s$^{-1}$. 
Since the I$_2$-cell RV technique only gives relative RVs, 
we roughly estimated the absolute stellar RV with respect to the barycenter of the solar system
from the position of H-$\alpha$ line as $\sim -44$ km s$^{-1}$.

%%%%%%%%%%%%%%%%%%%%%%%%%%%%%%%%%%%%%%%%%%%%%%%%%%%%%%%%%%%%%%%%%%%%%%
\begin{table}[tb]%[htb]
\caption{Radial velocities measured with Subaru/HDS. 
An arbitrary RV offset is subtracted in the data. 
}\label{tab:rv}
\begin{center}
\begin{tabular}{rrr}
\hline
Time [BJD (TDB)]  & Relative RV [m~s$^{-1}$] & Error [m~s$^{-1}$]\\
\hline\hline
2456473.12721 & -14.85 & 4.16\\
2456473.13152 & -13.02 & 4.23\\
2456473.13583 & -15.11 & 4.94\\
2456475.11812 & -4.70 & 4.26\\
2456475.12243 & -3.88 & 4.47\\
2456475.12677 & -9.69 & 4.69\\
2456475.13109 & -4.92 & 4.63\\
2456851.75827 & 22.54 & 3.93\\
2456851.96176 & -0.91 & 3.61\\
2456851.96955 & -10.11 & 3.80\\
2456852.12273 & -34.81 & 3.58\\
2456853.06078 & -5.28 & 5.05\\
\hline
\end{tabular}
\end{center}
\end{table}
%%%%%%%%%%%%%%%%%%%%%%%%%%%%%%%%%%%%%%%%%%%%%%%%%%%%%%%%%%%%%%%%%%%%%%

%%%%%%%%%%%%%%%%%%%%%%%%%%%%%%%%%%%%%%%%%%%%%%%%%%%%%%%%%
\subsection{Estimate for Spectroscopic Parameters}
Making use of the same template spectrum used for the RV analysis, we determined
the basic spectroscopic parameters. The procedure here to estimate the stellar parameters 
(e.g., stellar mass and rotational velocity $V\sin I_s$) is somewhat similar to that 
described in \citet{2012ApJ...756...66H, 2014ApJ...783....9H}:
we first measure the equivalent widths for all the available Fe I and II lines and estimate 
the stellar atmospheric parameters (the effective temperature $T_\mathrm{eff}$, 
surface gravity $\log g$, metallicity [Fe/H], micro-turbulent velocity $\xi$)
from the excitation and ionization equilibrium conditions \citep{2002PASJ...54..451T, 2005PASJ...57...27T}. 
Creating a theoretical spectrum with the ATLAS09 model for the best-fit
atmospheric parameters, we then optimized the projected rotational velocity $V\sin I_s$ 
by convolving that model spectrum with the IP and rotational plus macroturbulence 
broadening kernel for the radial-tangential model \citep{2005oasp.book.....G}. 
As for the macroturbulence, we adopt the macroturbulent velocity of 
$\zeta_{\rm RT}=1.7\pm 1.0$ km s$^{-1}$ based on the empirical relation between 
$T_\mathrm{eff}$ and $\zeta_{\rm RT}$ by \citet{2005ApJS..159..141V}.

%%%%%%%%%%%%%%%%%%%%%%%%%%%%%%%%%%%%%%%%%%%%%%%%%%%%%%%%%%%%%%%%%%%%%%
\begin{table}[tb]
\begin{center}
\caption{Stellar Parameters of KOI-977A}\label{hyo2}
%\begin{tabular}{lcl}
\begin{tabular}{lr}
\hline
Parameter& Value\\\hline\hline
(A) {\it Spectroscopy}&\\
$T_\mathrm{eff}$  (K)& 4275 $\pm$ 48 \\ %& \citet{2005PASJ...57...27T}\\
$\log g$ (dex) & $1.973\pm0.175$ \\ %& \citet{2005PASJ...57...27T}\\
$[\mathrm{Fe/H}]$ (dex) & $0.07\pm 0.09$ \\ %& \citet{2005PASJ...57...27T}\\
$\xi$ (km s$^{-1}$) & $1.13\pm 0.20$ \\ %& \citet{2005PASJ...57...27T}\\
$V\sin I_s$ (km s$^{-1}$) & $4.22\pm 0.27$ \\ %& \\
$M_\star$ ($M_\odot$) & $1.70_{-0.56}^{+1.27}$ \\ % & Y$^2$ isochrone\\
$R_\star$ ($R_\odot$) & $22.2_{-5.7}^{+13.1}$\\
%age (Gyr) & $2.1_{-1.71}^{+5.7}$ \\ %& Y$^2$ isochrone\\
\hline
(B) {\it Asteroseismology + Spectroscopy }&\\ 
$T_\mathrm{eff}$  (K)& 4280 $\pm$ 16 \\ 
$\log g$ (dex) & $1.876 \pm 0.004$ \\
$[\mathrm{Fe/H}]$ (dex) & $0.09\pm 0.03$ \\ 
$\log(L/L_\odot)$  &  $2.22 \pm 0.01$ \\
$M_\star$ ($M_\odot$) & $1.51 \pm 0.04$ \\ 
$R_\star$ ($R_\odot$) & $23.5 \pm 0.2$\\
${\rm Age}$ (Gyrs)    & $3.25 \pm 0.25$ \\ \hline
\end{tabular}
\end{center}
\end{table}
%%%%%%%%%%%%%%%%%%%%%%%%%%%%%%%%%%%%%%%%%%%%%%%%%%%%%%%%%%%%%%%%%%%%%%
We next convert the atmospheric parameters into the stellar mass $M_\star$ and radius $R_\star$
employing the Yonsei-Yale (Y$_2$) isochrone model \citep{2001ApJS..136..417Y}. 
We implement a Monte Carlo simulation, varying the atmospheric
parameters: we randomly generate $T_\mathrm{eff}$, $\log g$, and [Fe/H] based on their 
spectroscopic errors assuming Gaussian distributions, and each set of the parameters 
is converted into the mass $M_\star$ and radius $R_\star$. The resulting distributions 
give the estimates and their uncertainties for $M_\star$ and $R_\star$. 
Thus derived spectroscopic parameters are summarized in Table \ref{hyo2} (A).

%%%%%%%%%%%%%%%%%%%%%%%%%%%%%%%%%%%%%%%%%%%%%%%%%%%%%%%%%
\subsection{Fitting RV Data}
The RV data obtained in \S \ref{s:rv} indicate that the system does
not contain within a few days a massive companion, which leads to an RV amplitude of 
$\gtrsim 100$ m s$^{-1}$. 
%This fact rules out the possibility that the observed EV in the {\it Kepler}
%data is caused by a binary star and the light-curve is diluted by a background/foreground 
%object so that the amplitude of EV is relatively small. 
We here fit the RV data alone in order to put a rough constraint on the mass
of the possible companion to KOI-977.  
The $\chi^2$ statistics for the RV fit is 
%%%%%%%%%%%%%%%%%
\begin{eqnarray}
\chi_\mathrm{RV}^2 =\sum_{i}\frac{(v_\mathrm{obs}^{(i)}-v_\mathrm{model}^{(i)})^2}{\sigma_\mathrm{RV}^{(i)2}},
\end{eqnarray}
%%%%%%%%%%%%%%%%%
where $v_\mathrm{obs}^{(i)}$ and $\sigma_\mathrm{RV}^{(i)}$ 
are the $i$-th observed RV and its error. As usual, RVs are modeled as a function of 
the true anomaly $f$ as
%%%%%%%%%%%%%%%%%
\begin{eqnarray}
\label{eq:RVmodel}
v_\mathrm{model}=K\{\cos(f+\omega)+e\cos\omega\}+\gamma, 
\end{eqnarray}
%%%%%%%%%%%%%%%%%
where $K$, $e$, $\omega$, and $\gamma$ are the RV semi-amplitude, 
orbital eccentricity, argument of periastron, and RV offset of our dataset, respectively. 
Allowing $K$, $e\cos\omega$, $e\sin\omega$, and $\gamma$ to be free,
we implement Markov Chain Monte Carlo (MCMC) algorithm to estimate
the posterior distributions for those fitting parameters from the observed RVs. 
The median and 15.87 and 84.13 percentiles of the marginalized posterior distribution
are taken to be the best-fit value and its $\pm 1\sigma$ uncertainty. 

%%%%%%%%%%%%%%%%%%%
\begin{figure}[t]
\begin{center}
\includegraphics[width=9cm,clip]{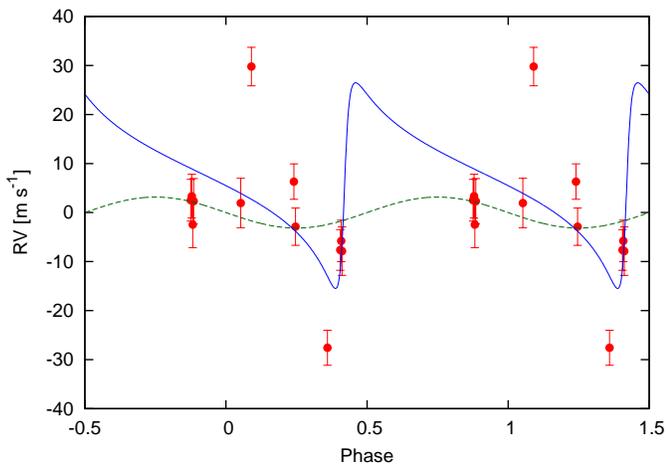} 
\caption{RV data obtained by Subaru/HDS, folded with the orbital period of KOI-977.01. 
The blue (solid) and green (dashed) lines indicate the best Keplerian fits to the observed RVs
with a noncircular and circular model, respectively. 
}\label{fig:rvonly}
\end{center}
\end{figure}
%%%%%%%%%%%%%%%%%%%
As a result, we obtain $K=21.0_{-4.1}^{+4.2}$ m s$^{-1}$, $e\cos\omega=-0.0829_{-0.0057}^{+0.0069}$,
and $e\sin\omega=-0.750_{-0.048}^{+0.058}$, and the best-fit model is plotted by the blue
solid line along with the RV data points (red) in Figure \ref{fig:rvonly} as a function of orbital phase. 
Our result for the RV fit indicates that the orbit is highly eccentric ($e\sim 0.75$), which is remarkably 
inconsistent with the phase-folded LC light-curve for KOI-977.01 shown in Figure \ref{fig:phasebest}. 
Figure \ref{fig:phasebest} suggests that the local minima of the modulation 
by ellipsoidal variation occur at $\phi\sim 0.0$ and $\phi\sim 0.5$, an indication that $e$ is small. 
%Assuming the stellar mass of $M_\star=1.7M_\odot$, 
%the projected mass of KOI-977.01 is estimated as $M_p\sin I_o\simeq 0.11M_J$. 
We also fit the observed RVs with a circular orbit ($e=0$). 
The resulting RV semi-amplitude is $K=3.16_{-1.86}^{+1.85}$ m s$^{-1}$ 
(the green dashed line in Figure \ref{fig:rvonly}),
which is consistent with a non-detection within $2\sigma$. 
There are a few RV data points that strongly disagree with both eccentric and circular models,
but these could be ascribed to large RV jitters ($10-20$ m s$^{-1}$) by stellar pulsation of 
the red giant with a timescale of $\sim 1$ day.
The implication of the small overall RV signal will be discussed in Section \ref{s:reduction}.

%%%%%%%%%%%%%%%%%%%%%%%%%%%%%%%%%%%%%%%%%%%%%%%%%%%%%%%%%%%%%%%%%%%%%%
\section{Estimate for Stellar Parameters from Asteroseismology \label{s:seismology}}\label{s:seismology}
In this section, we also attempt to obtain an estimate for stellar parameters
from asteroseismology, using {\it Kepler} public light-curve. 

\subsection{Asteroseismology of Solar-like Pulsators} \label{sec:seism:intro}
Solar-like pulsators are cool (i.e., $\lesssim 7000$ K) and low mass (i.e., $\approx 0.8 - 2.5 M_\odot$) stars, mostly characterized by an outer convective layer which excites pulsation modes. 
Asteroseismology, the science that studies these pulsations, enables high-precision estimations of stellar fundamental parameters \citep{Lebreton2009} (e.g., uncertainties of a few percent for their mass and radius).

For a spherically symmetric star, each mode is characterized
by three quantum numbers: the angular degree $l$, the azimuthal
order $m$ ($-l \le m \le +l$), and the radial order $n$. The degree $l$ corresponds to the number of nodal surface lines, while the
azimuthal order $m$ specifies the surface pattern of the eigenfunction, with $|m|$ being the number of longitude lines among the $l$ nodal
surface lines. The radial order $n$ corresponds to the number of nodal
surfaces along the radius.  In the absence of rotation, the radial
eigenfunction and the frequency of each mode are independent of $m$ and
show a $(2l+1)$-fold degeneracy.  Frequencies of high order, acoustic
(or p-) modes of the same low degree $(n \gg l \sim 1)$ are regularly
spaced and separated on average by a frequency spacing
$\Delta\nu \propto \sqrt{\rho}$, sensitive to the mean stellar density $\rho$. Furthermore, mode amplitudes arise from a competition between damping and excitation that is a function of the frequency. The competition actually makes amplitudes follow a bell shaped function that peaks at $\nu_{\rm max}$, the frequency at the maximum amplitude. 

When combined with the stellar effective temperature $T_{\rm eff}$, $\nu_{\rm max}$ and $\Delta\nu$ enable 
us to evaluate the stellar mass and radius using the so-called \emph{scaling relations} \citep[e.g.,][]{Kjeldsen1995, Huber2011},
\begin{subequations}
	\begin{equation} \label{eq:scalrelR}
		R_\star = \left(\frac{\nu_{\rm max}}{\nu_{\rm max, \odot}} \right) \left(\frac{\Delta\nu}{\Delta\nu_\odot} \right)^{-2}  \left( \frac{T_{\rm eff}}{T_{\rm eff, \odot}} \right)^{1/2} \text{ and,}
	\end{equation}
	\begin{equation} \label{eq:scalrelM}
		M_\star = \left(\frac{\nu_{\rm max}}{\nu_{\rm max, \odot}} \right)^{3} \left(\frac{\Delta\nu}{\Delta\nu_\odot} \right)^{-4} \left(\frac{T_{\rm eff}}{T_{\rm eff, \odot}} \right)^{3/2}.
	\end{equation}
\end{subequations}
In these empirical relations, %\textbf{
the Sun acts as a reference star and we adopted solar $T_{\rm eff, \odot} = 5777 K$ \citep[e.g.,][]{Kjeldsen1995}, $\Delta\nu_\odot = 135.1 \pm 0.1 \, \mu$Hz and $\nu_{\rm max, \odot} = 3150 \pm 50 \, \mu$Hz. The seismic values are obtained using the same method as for KOI-977 (see Section \ref{sec:seism:koi977}) but using photometry from the VIRGO instrument aboard SoHo \citep{1997SoPh..170....1F}. Note that $\nu_{\rm max, \odot}$ slightly differs from that in \cite{Huber2011} who analysed the same data, but stays consistent at $1\sigma$.

While for main sequence stars we observe only acoustic modes, subgiants and red giants also show a rich spectrum of so-called \emph{mixed modes} \cite[e.g.,][]{Beck2011Science, Bedding2011Nature, Benomar2013a, 2013Sci...342..331H}. They have a dual nature, behaving as acoustic modes in the stellar envelope and as gravity modes in the core.  They are therefore detectable at the stellar surface while also probing deep stellar layers, potentially providing stringent constraints on the stellar evolution stage. However, for latest evolution stages (e.g., tip of the RGB and AGB phase), the observation duration required to resolve the mixed modes is of several years and these could be hard to measure even with the exquisite \emph{Kepler} data.

%such as,
%%%%%%%%%%%%%%%%%%%%%%%%%%%%
%\begin{equation}
%\nu(n,l) = \Delta\nu \left(n+{{l}\over{2}}+\alpha \right) 
%+ \varepsilon_{n,l},
%	\label{eq:2}
%\end{equation}
%%%%%%%%%%%%%%%%%%%%%%%%%%%
%where $\alpha$ is a constant of order unity, and $\varepsilon_{n,l}$ is
%a small correction.  Note also that $\Delta\nu$ is proportional to the mean stellar density. 

\subsection{\emph{Kepler} Data Processing of KOI-977}
We downloaded the light-curve data from the {\it Kepler} MAST archive, and
used the PDC-SAP flux, for which unphysical trends (e.g., instrumental artifacts) 
are designed to be removed \citep{2012PASP..124.1000S, 2012PASP..124..985S}. 
Among the public light-curves for KOI-977, short-cadence (SC) 
data were available for Q8 only, which involve 15 transits of KOI-977.01. 
On the other hand, long-cadence (LC) data were available for all quarters (Q0 to Q17). 

\begin{figure}[t]
\begin{center}
\includegraphics[width=6.7cm,angle=90, clip]{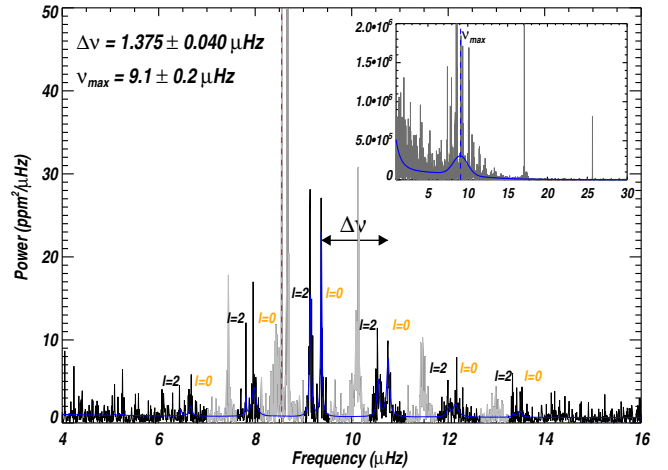} 
\caption{Power spectrum of KOI-977. \textbf{Main figure.} Pulsation modes of degree $l=0$, $l=1$ and $l=2$ are seen. The best Lorentzian fit of $l=0$ and $l=2$ is indicated in blue (solid-line). The dense spectra of modes of degree $l=1$ (grey areas) do not allow us to disentangle them. They are ignored during the mode analysis. The vertical red line indicates the fundamental frequency due to periodic eclipses. \textbf{Inset.}  Spectrum over a frequency range that includes the two first orbital period harmonics. The blue line corresponds to the best gaussian fit of the mode envelope.}
\label{fig:psf}
\end{center}
\end{figure}

LC data are first appended and corrected from jumps between successive quarters in order to obtain a continuous smooth time series. Furthermore, we use the mean flux to convert the time series into ppm. The power spectrum (Figure \ref{fig:psf}) is computed by means of the Lomb-Scargle periodogram, suitable for unevenly sampled data \citep{Scargle1982}. The power spectrum shows an excess of power around $\nu \approx 10 \mu$Hz due to stellar pulsations, which is compatible with an evolved red giant. We applied a similar procedure using the raw time series, but using a method similar to \cite{Garcia2011} for removing artifacts and applying a median high pass filter to remove long-term variations (longer than 11.57 days). The power spectrum using PDC-SAP or raw data are nearly identical, indicating that the stellar signal is not affected by the data processing method. Finally, we used SC data to investigate the high frequency pulsation due to hypothetical companions, but no evidence was found. Thus, if KOI-977 is a part of a multiple stellar system, none of its companions show measurable pulsations. 
%%%%%%%%%%%%%%%%%%%
%%%%%%%%%%%%%%%%%%% 

Figure \ref{fig:psf} shows a narrow peak at $8.548$ $\mu$Hz and its harmonics, caused by the periodic eclipses. 
These peaks may pollute the stellar signal. However, most of their power is confined within only 2 or 3 bins, 
we decided to simply ignore the few affected data points during the following analysis.

\subsection{KOI-977 Seismic Analysis}\label{sec:seism:koi977}
The stellar signal analysis is threefold. Firstly, we obtain an approximate mass and radius of the star using the scaling relations (Equations (\ref{eq:scalrelR}) and (\ref{eq:scalrelM}) in Section \ref{sec:seism:intro}).
 %The scaling relations are empirical relations relying on the frequency at maximum amplitude $\nu_{max}$ as well as the frequency spacing $\Delta\nu$. 
The frequency $\nu_{\rm max}$ is obtained by fitting a Gaussian on the excess of power due to the modes \citep[see e.g.,][]{Mathur2010, Benomar2012}, while $\Delta\nu$ is obtained by performing the power spectrum autocorrelation, and by ignoring $l=1$ mixed modes (grey areas in Figure \ref{fig:psf}). We obtain $\nu_{\rm max}=9.1 \pm 0.2\mu$Hz and $\Delta\nu=1.375 \pm 0.040\mu$Hz, which lead to $M_\star=1.49 \pm 0.11M_\odot$ and $R_\star=24.3 \pm 0.8R_\odot$.

These first estimates may suffer from systematics because the scaling relation is calibrated using the Sun for reference, whereas KOI-977 is a much evolved star. Thus, to confirm the mass and radius, we attempted to identify a stellar structure model that best match both individual pulsation frequencies and spectroscopic constraints. 

Each individual acoustic mode can be seen as a stochastically excited oscillator whose limit spectrum is a Lorentzian. Therefore the Lorentzian is often used to extract the pulsation mode characteristics (such as the mode frequency) in the power spectrum \cite[e.g.,][]{Appourchaux1998, Chaplin1998, Appourchaux2008}. Here, the fit was performed using an MCMC approach, ensuring both precision and accuracy \citep{Benomar2008, Benomar2009, Benomar2009b, Handberg2011, Benomar2014}. 
Note that although $l=1$ mixed modes are clearly visible in the power spectrum, their nearly identical frequencies and high number prevent us to resolve them individually. Thus only modes of degrees $l=0$ and $l=2$ are fitted with Lorentzians in this analysis. We detected a total of 12 modes\footnote{The modes of degree $l=0$ and $l=2$ over 6 radial orders.} and the best corresponding fit is shown in Figure \ref{fig:psf}. Note that we could not measure the internal rotation and the stellar inclination, due to the small number of measured non-radial modes.

Finally, stellar models that simultaneously match spectroscopic and seismic observables are found using the `astero' module of
the Modules for Experiments in Stellar Astrophysics (MESA) evolutionary code \citep{Paxton2010}.  Stellar models are calculated assuming a fixed mixing length parameter $\alpha_{\rm MLT}=2.0$ and an initial hydrogen abundance $X=0.7$. The opacities are
calculated using the OPAL opacities \cite{Iglesias1996} and the solar composition from \cite{Asplund2009}. 
We use NACRE Nuclear reactions \citep{Angulo1999} with updated $^{14}{\rm N}(p, \gamma)^{15}{\rm O}$ 
and $^{12}{\rm C}(\alpha, \gamma)^{16}{\rm O}$  reactions \citep{Kunz2002, Formicola2004}. We decided not to implement overshoot and diffusion. %\textbf{
Because these processes enhance the mixing of the chemical elements within a star, neglecting them may significantly bias the stellar age. Finally note that only mass $M_\star$, metallicity [Fe/H] and age are treated as free parameters.

Eigenfrequencies are calculated assuming adiabaticity and using {\tt ADIPLS} \citep{JCD2008b}. We did not apply the surface correction.  The search for the best model involves a simplex minimization approach \citep{Simplex} using the $\chi^2$ criteria. Uncertainties are estimated by evaluating the $\chi^2$ for solutions surrounding the best model and by weighing the model parameters with the likelihood $\propto \exp(-\chi^2/2)$. Note that the reported uncertainties do not account of hardly quantifiable uncertainties due to the used physics. 

\begin{figure}[t]
\begin{center}
\includegraphics[width=9.2cm]{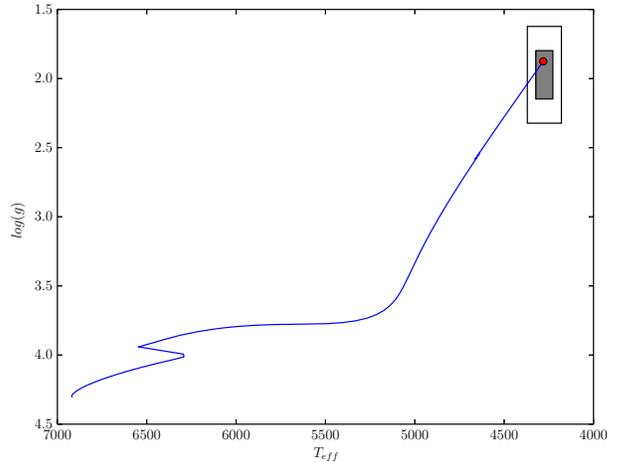} 
\caption{$\log g -T_{\rm eff}$ diagram of KOI-977. The blue line indicates the theoretical evolution of a $M=1.51M_\odot$ star. The model that best matches seismic and spectroscopic observables is indicated in red. Uncertainties from spectroscopy at $1\sigma$ (inner grey box) and $2\sigma$ (outer box) are also shown.}
\label{fig:loggTeffdiag}
\end{center}
\end{figure}

Figure \ref{fig:loggTeffdiag} shows the $\log g-T_{\rm eff}$ diagram along the evolution of the best matching model 
%\textbf{
and Table \ref{hyo2}B gives the fundamental parameters of the model. 
The model describes a star of $M_\star=1.51 \pm 0.04 M_\odot$ and $R_\star=23.5 \pm 0.2 R_\odot$, compatible with our first seismic estimate and with the isochrone fitting (involving spectroscopic constraints alone). %\textbf{
The model%} 
has exhausted its Hydrogen in the core and seems to have ignited Helium fusion reaction, which indicates that it is in the red clump. To verify this, we attempted to measure the so-called \emph{observed period spacing}  of $l=1$ mixed modes $\Delta P_{1}$, using the same approach as \cite{Bedding2011Nature}. This quantity is sensitive to the core structure and is a stringent indicator of the evolutionary stage of stars. Note that here we use a power spectrum, oversampled by a factor 10, to improve the precision of our measurement which relies on the autocorrelation of the power spectrum. We obtain $\Delta P_1 \approx 200 - 300$ sec, depending on which cluster of $l=1$ mixed modes we considered\footnote{A cluster of mixed modes corresponds to an ensemble of mixed modes with very close frequency and of high amplitude. In Figure \ref{fig:psf}, each cluster of mixed modes is shown in grey and correspond to an almost undistinguishable forest of $l=1$ mixed modes.}. According to \cite{Bedding2011Nature}, this is typical of red clump stars or of secondary clump stars\footnote{Stars too massive to have undergone a helium flash.}, while stars that did not ignited Helium fusion reaction (red giant branch) have a period spacing approximately five times smaller.

%%%%%%%%%%%%%%%%%%%
%%%%%%%%%%%%%%%%%%%

%%%%%%%%%%%%%%%%%%%%%%%%%%%%%%%%%%%%%%%%%%%%%%%%%%%%%%%%%%%%%%%%%%%%%%
\section{Model and Analysis of {\it Kepler} Photometric Data \label{s:model}}\label{s:model}
%%%%%%%%%%%%%%%%%%%%%%%%%%%%%%%%%%%%%%%%%%%%%%%%%%%%%%%%%
\subsection{Reduction of the {\it Kepler} Light-curves \label{s:reduction}}
The analyses with spectroscopy and asteroseismology both indicate that
the stellar radius of KOI-977 is no less than $\sim 20R_\odot$, which corresponds
to $\sim 0.093$ AU. Using Kepler's third law, however, we obtain the estimated semi-major 
axis of $\sim 0.027$ AU from KOI-977.01's orbital period and an assumed
stellar mass of $M_\star=1.51 M_\odot$. Therefore, if the pulsating giant KOI-977 is 
orbited by KOI-977.01, it has to be inside the star. 
In addition, the transit depth of KOI-977.01 ($\sim 0.08\%$) 
and host star's radius of $23.5R_\odot$ translate 
into companion's radius of $\sim 0.7R_\odot$, which corresponds to a star rather than a planet. 
A star of this size would cause an RV variation of $K\sim 80$ km s$^{-1}$, which strongly 
contradicts the observed RV amplitude. 
These facts made us conclude that
KOI-977.01 is likely a false positive, meaning that the eclipsing binary (or transiting planetary 
system) is different from the most luminous giant for which we estimated
stellar parameters and measured RVs. According to the WIYN 
speckle images on CFOP website\footnote{https://cfop.ipac.caltech.edu/}, 
KOI-977 may have a possible 
companion at a separation of $\sim 0.34^{\prime\prime}$ with magnitude differences of 
$\Delta m\sim 3.94$ and $\Delta m\sim 4.06$ at 692 nm and 880 nm, respectively.
Therefore, a natural explanation is that this faint companion is an eclipsing binary
causing periodic depressions diluted by the red giant so that the depth of eclipse 
is comparable to that of a planetary transit ($\sim 10^{-3}$).

In the following discussion, we assume that KOI-977 is comprised of a red giant
and an eclipsing binary, and search for a possible solution to the observed
properties (i.e., transits and phase-curve variation). We name the pulsating giant
and the brighter star in the eclipsing binary as KOI-977A  and KOI-977B, respectively, 
and call the fainter one as KOI-977.01, which is orbiting KOI-977B. 
In order to extract system parameters from transits and phase-curve variations, 
we reduce the same public light-curves for KOI-977 as those used for asteroseismology
and phase-fold them by the procedure below. 
To avoid removing possible astrophysical signals, we decided not to further detrend
the PDC-SAP light-curves, and simply normalize them by diving the PDC-SAP flux
by the median flux value for each quarter. Combining all the LC light-curves
from Q0 to Q17, we phase-fold the combined light-curve with the orbital 
period and transit ephemeris reported by the official {\it Kepler} team. 
We later take into account the impact of incorrect transit ephemeris.

Next we bin the folded LC light-curve.
The number of bins is fixed at $65$, which was adopted so that each bin approximately 
covers the time span of LC data point ($\sim 30$ minutes). 
We noticed that adopting a larger number of bins leads to a noisier binned light-curve, 
supposedly due to a contamination of the quasi-periodic stellar pulsation. 
The binned flux for each bin is computed as the median flux value of that bin. 
As for the flux error for each bin, we first compute the root-mean-square (RMS) of the flux
residuals from the median flux, and divide the RMS by the square root of the number of
points within the bin. Thus derived flux error, however, becomes much larger 
($\gtrsim 4.5\times10^{-5}$) than the apparent dispersion of the binned flux. 
The quasi-periodic flux modulation from KOI-977A results in an apparent flux dispersion
in the folded light-curve, but its impact is likely alleviated when the light-curve is binned. 
Hence, we decide to scale the binned flux error when we fit the light-curves; the error bars
are iteratively scaled so that $\chi^2$ for the binned LC light-curve becomes equivalent to 
the degree of freedom for LC data fitting (the reduced $\chi^2$ of unity). In the following analysis, 
this scaling is always applied during the $\chi^2$ optimization. 

%%%%%%%%%%%%%%%%%%%
%\begin{figure}[t]
%\begin{center}
%\includegraphics[width=9cm,clip]{lccurve_koi977.eps} 
%\caption{Binned LC data folded with KOI-977.01's period. 
%}\label{fig:lccurve}
%\end{center}
%\end{figure}
%%%%%%%%%%%%%%%%%%%
As shown in Figure \ref{fig:phasebest}, the binned LC light-curve shows a clear pattern
of periodic flux modulation. This bimodal modulation that is synchronous with the orbital 
period of KOI-977.01 (two peaks within one period) is representative of a stellar ellipsoidal variation, 
but one exception is that the flux peaks at an orbital phase of $\phi\sim 0.75$. 
The flux modulation due to the Doppler boosting shows a flux peak
around the orbital phase of $\phi\sim 0.25$. 
There are some possibilities that lead to such phase-curve behaviours;
when the transiting object is a white dwarf, for instance, which is smaller in radius 
but as massive as a normal star, the phase-curve modulation may
peak at the observed phase. In this case, the depression at $\phi\sim 0.0$
in Figure \ref{fig:phasebest} corresponds to a ``secondary eclipse" of KOI-977.01. 
One issue concerning this possibility is, however, that the phase-curve modulation due 
to the ellipsoidal variation should have a higher flux at the secondary eclipse ($\phi\sim 0.0$)
than the flux around $\phi\sim 0.5$ in the absence of transits/eclipses \citep{2012ApJ...751..112J}, 
but the reverse is true for Figure \ref{fig:phasebest}, suggesting that this scenario is very unlikely.

Another possibility is a ``phase-shift" of the
KOI-977.01's flux maximum from the superior conjunction. Because of the high luminosity 
of KOI-977B and close-in orbit, the dayside surface of KOI-977.01 is highly heated. 
There have been reports on 
the possibility that the emission/reflected light from close-in planets have shifted maxima, 
either eastward or westward, from the substellar point 
\citep[e.g.,][]{2007Natur.447..183K, 2013ApJ...776L..25D, 2014arXiv1407.2245E}. 
Whether KOI-977.01 is a star or planet (brown dwarf), the high irradiation to 
KOI-977.01 may lead to a temperature inhomogeneity and the location 
with the highest intensity is not necessarily the substellar point. 
%We here ascribe the observed phase-curve behaviour (i.e., a higher peak at $\phi\sim 0.75$) 
%to the phase-shift of the maximum local intensity, and this phase-delay is incorporated 
%into our modeling of the light-curve. 
Below, as an ``empirical" treatment for the asymmetric flux maximum, 
we introduce a phase-delay following \citet{2013ApJ...776L..25D}. 
We will revisit this issue in Section \ref{s:discussion}.

%%%%%%%%%%%%%%%%%%%%%%%%%%%%%%%%%%%%%%%%%%%%%%%%%%%%%%%%%
\subsection{Light-curve Models \label{s:lcmodel}}
When the objects are very close to each other, a sinusoidal model does not give a good 
fit to the stellar ellipsoidal variation \citep[e.g.,][]{2007ApJ...661.1129V}. 
Our model for the binned LC light-curve is based on the EVIL-MC model 
\citep{2012ApJ...751..112J}, which is easy to compute and applicable to 
a wide range of the mass ratio and semi-major axis 
\citep[see Figure 3 in][]{2012ApJ...751..112J}. 
In the EVIL-MC model, the relative intensity at wavelength 
$\lambda$ at position $\mathbf{R_\star}$ on the ellipsoidally distorted stellar surface is 
expressed as 
%%%%%%%%%%%%%%%%%
\begin{eqnarray}
\label{eq:1}
I(f, \lambda, \mathbf{R_\star}) &\propto& \left\{1+\beta
\frac{x}{1-\exp(-x)}(\mathbf{R_0}\cdot\mathbf{\delta\Gamma_0}
-\mathbf{R_\star}\cdot\mathbf{\delta\Gamma})\right\}\nonumber\\
&&\left[1+\frac{x}{1-\exp(-x)}\frac{K}{c}\{\cos(f+\omega)+e\cos\omega\}\right]
\nonumber\\
&&\{1-u_1(1-\mu)-u_2(1-\mu)^2\},
\end{eqnarray}
%%%%%%%%%%%%%%%%%
where $\beta$ is the gravity darkening exponent, $c$ is the speed of light, 
$x=hc/k_B\lambda T_\mathrm{eff}$, 
$\mathbf{R_0}$ is the position vector which
is normal to both star's spin vector and planet's position vector, 
$\mu$ is the cosine of the angle between our line-of-sight 
and normal vector to the local stellar surface, and $u_1$, $u_2$ are the quadratic 
limb-darkening coefficients, respectively. 
Note that $\mathbf{R_\star}$ and $\mathbf{R_0}$ are both normalized by $|\mathbf{R_0}|$
(i.e., $\mathbf{R_0}$ is a unit vector) in Equation (\ref{eq:1}). 
The vector $\mathbf{\delta\Gamma}$ 
represents the deviation of the normalized gravity vector from the unperturbed 
gravity vector at $\mathbf{R_\star}$, and $\mathbf{\delta\Gamma_0}$ is that at $\mathbf{R_0}$ 
\citep[see][for the expressions of these quantities]{2012ApJ...751..112J}. 
The observed flux for a given orbital phase is obtained by integrating Equation (\ref{eq:1}) 
over wavelength $\lambda$ (through {\it Kepler's} band) and visible stellar hemisphere.

We employ the EVIL-MC model in the discussion below, 
but we slightly revise the model on the following three aspects. 
\begin{enumerate}
\item To allow for a non-zero orbital eccentricity, we allow for the variable distance
between the planet and star, and multiply $a$ ($=a/R$ here) in Equations (6) and (8) 
in \citet{2012ApJ...751..112J} by
\begin{eqnarray}
\displaystyle \frac{1-e^2}{1+e\cos f}. 
\end{eqnarray}
\item In the EVIL-MC model, the emission/reflected light from the planet is modeled by a simple 
sinusoidal model and the total flux from the planet $F_p$ is expressed as a function $\phi$ as
\begin{eqnarray}
F_p = F_0 - F_1\cos(2\pi\phi),
\end{eqnarray}
where $F_0$ is the flux offset arising from the homogeneous surface emission, 
$F_1$ is the flux variation amplitude of planet's emission/reflected light.
Here we take into account the effect of non-zero eccentricity and 
introduce an empirical phase-delay of KOI-977.01's flux maximum from the 
superior conjunction as stated in \S\ref{s:reduction}:
\begin{eqnarray}
\label{eq:planetlight}
F_p = F_0 - F_1\left(\frac{1+e\cos f}{1-e\sin\omega}\right)^2\sin(f+\omega-2\pi\Delta\phi),
\end{eqnarray}
where $\Delta\phi$ is the phase-shift of the brightest part of KOI-977.01. 
%Note that KOI-977.01 could be a star rather than a planet, but due to the high irradiation 
%from KOI-977B, the local intensity is more or less inhomogeneous on KOI-977.01's surface 
%and the irradiated side should be brighter than the night side. 
%Thus, the expression above is also applicable to the present case. 
The fluxes $F_0$ and $F_1$ are normalized by KOI-977B's flux at $\phi=0.5$, 
and $F_p$ is added to KOI-977B's flux obtained by integrating Equation (\ref{eq:1}). 
\item We incorporate the light-curve models during the primary transit and secondary eclipse
phases; here we simply call the occultation of KOI-977B by KOI-977.01 as ``transit" 
($\phi\sim 0.0$) and the occultation of KOI-977.01 by KOI-977B as ``secondary eclipse" 
($\phi\sim 0.5$). 
The model for phase-curve variations (described above) is multiplied by the analytic
transit/eclipse model by \citet{2009ApJ...690....1O}, which is also applicable to eclipsing
binaries. Thus, the parameters in phase-curve variations and transit/eclipse 
(e.g., $a/R$, $e\cos\omega$, and $e\sin \omega$) are simultaneously determined 
in fitting the observed light-curve. 
\end{enumerate}
In addition to the above revisions, we introduce a dilution factor $D$ for KOI-977, which
is defined as the flux ratio of KOI-977A to KOI-977B in the absence of KOI-977.01. 
The precise dilution factor for the {\it Kepler} band is not known, but we here assume 
$D=37$ based on the contrasts at 692 nm and 880 nm. This dilution factor is added
to the modeled (normalized) phase-curve for KOI-977B. We later try different values 
of $D$ and see the difference in the derived parameters.

Binned LC data are useful in the sense that flux variations originating from
the stellar pulsation and/or activity of KOI-977A is cancelled out and we can extract 
the light-curve modulation synchronous with KOI-977.01's orbital period. 
However, it is difficult with LC data alone to precisely estimate the transit model
parameters (e.g., the mid-transit time $T_c$ and limb-darkening parameters)
and discuss a possibility of TTVs. Thus, we also make use of 
the available SC data, for which fifteen transits are simultaneously fitted. 
Inclusion of the SC data in the global modeling also improves the accuracy of the fit
and helps to break the degeneracy between correlated system parameters (e.g., 
the semi-major axis scaled by KOI-977B's radius $a/R_\mathrm{B}$,
orbital inclination $I_o$, and the mass ratio $q$).  

%%%%%%%%%%%%%%%%%%%
\begin{figure}[t]
\begin{center}
\includegraphics[width=9cm,clip]{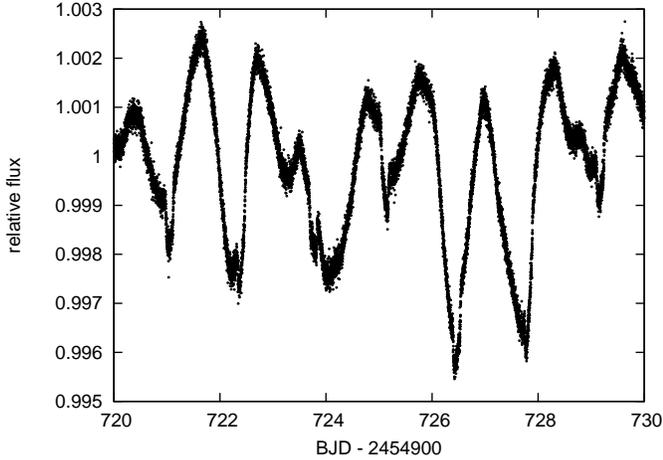} 
\caption{Part of SC light-curve (Q8) of KOI-977. A time-variable modulation due to 
stellar pulsation is visible which is larger in amplitude than the transit depth. 
}\label{fig:sc}
\end{center}
\end{figure}
%%%%%%%%%%%%%%%%%%%
As Figure \ref{fig:sc} shows, the unfolded light-curve of KOI-977 has a strong modulation 
due to KOI-977A's pulsation. 
We thus decide to cut the light-curve into fifteen segments, 
each of which covers a transit of KOI-977.01 along with a baseline
before and after the transit. We adopt the length of the baseline as $\sim 0.2$ day
from the predicted transit center toward both sides of the transit. 
Each light-curve segment is modeled as a polynomial 
function (describing the flux baseline), which is multiplied by the analytic transit
model by \citet{2009ApJ...690....1O}. The parameters relevant to the SC
light-curve are $a/R_\mathrm{B}$, $I_o$, $u_1$, $u_2$, 
$R_\mathrm{01}/R_\mathrm{B}$, $e$, $\omega$, $\Delta T_c^{(j)}$ (the small correction 
for the mid-transit time from {\it Kepler's} official ephemeris ($T_{c,0}=2455094.434370$) 
for segment $j$ of the SC data), and coefficients of polynomials. 
The order of polynomials is discussed in the next subsection.

%%%%%%%%%%%%%%%%%%%%%%%%%%%%%%%%%%%%%%%%%%%%%%%%%%%%%%%%%
\subsection{Fitting Light-curves\label{sec:fitcurve}}
We here fit the light-curves reduced in \S \ref{s:reduction}. 
In order to estimate system parameters, we simultaneously model and fit LC and SC light-curves. 
We employ the following $\chi^2$ statistics here: 
%%%%%%%%%%%%%%%%%
\begin{eqnarray}
\label{eq:chisq}
\chi^2 =\chi^2_\mathrm{LC} + \chi^2_\mathrm{SC},
\end{eqnarray}
%%%%%%%%%%%%%%%%%
where
%%%%%%%%%%%%%%%%%
\begin{eqnarray}
%\sum_{i}\frac{(v_\mathrm{obs}^{(i)}-v_\mathrm{model}^{(i)})^2}{\sigma_\mathrm{RV}^{(i)2}}
\label{eq:chisqlc}
\chi^2_\mathrm{LC}&=&
\sum_{i}\frac{(f_\mathrm{LC,obs}^{(i)}-f_\mathrm{LC, model}^{(i)})^2}{\sigma_\mathrm{LC}^{(i)2}},\\
\label{eq:chisqsc}
\chi^2_\mathrm{SC}&=&
\sum_{j}\sum_{i}\frac{(f_\mathrm{SC,obs}^{(j, i)}-f_\mathrm{SC, model}^{(j, i)})^2}{\sigma_\mathrm{SC}^{(j, i)2}},
\end{eqnarray}
%%%%%%%%%%%%%%%%%
where $f_\mathrm{LC,obs}^{(i)}$ and $\sigma_\mathrm{LC}^{(i)}$ 
are the $i$-th observed binned LC flux and its error, respectively. 
The $i$-th SC flux of the light-curve segment $j$ ($1\leq j\leq 15$) and its error are represented 
by $f_\mathrm{SC,obs}^{(j, i)}$ and $\sigma_\mathrm{SC}^{(j, i)}$, respectively. 
As stated in Section \ref{s:lcmodel}, $f_\mathrm{LC, model}$ is computed by
integrating Equation (\ref{eq:1}), which is multiplied by the transit/eclipse model,
with the addition of the planet flux $F_p$ and the dilution factor $D$. We employ the gravity darkening 
exponent of $\beta=0.095$ based on the table by \citet{1998A&AS..131..395C}
for $\log T_\mathrm{eff}=3.76$ and $\log g=4.4$. The RV semi-amplitude in Equation (\ref{eq:1})
is computed by 
%%%%%%%%%%%%%%%%%
\begin{eqnarray}
\label{eq:K}
K=212908.30\left(\frac{M_\mathrm{B}/M_\odot}{P_\mathrm{orb}/\mathrm{day}}\right)^{\frac{1}{3}}
\frac{q}{(1+q)^{\frac{2}{3}}}\frac{\sin I_o}{\sqrt{1-e^2}}~~~~(\mathrm{m~s}^{-1}),\nonumber\\
\end{eqnarray}
%%%%%%%%%%%%%%%%%
where $q$ is the mass ratio ($=M_\mathrm{01}/M_\mathrm{B}$) .

The adopted fitting parameters in our model for LC and SC light-curves are $a/R_\mathrm{B}$, 
$b$ (the transit impact parameter), $u_1+u_2$, $u_1-u_2$, $q$, $F_0$, $F_1$, $\Delta \phi$, $C$ 
(the overall normalization factor of the binned LC light-curve), $R_\mathrm{01}/R_{\rm B}$, $e\cos\omega$, 
$e\sin\omega$, $\Delta T_c^{(\mathrm{LC})}$ (the mid-transit time correction for LC data), 
$\Delta T_c^{(j)}$ ($1\leq j\leq 15$), 
and the coefficients of polynomials for SC light-curve segments. 
Note that the transit impact parameter $b$ is related to the orbital inclination $I_o$ by
%%%%%%%%%%%%%%%%%%%%%%
\begin{eqnarray}
\label{impact_parameter}
b=\frac{a\cos I_o}{R_\mathrm{B}}\left(\frac{1-e^2}{1+e\sin \omega}\right).
\end{eqnarray}
%%%%%%%%%%%%%%%%%%%%%%
Assuming that each SC light-curve baseline is expressed by an $n$-th order polynomial, the number of
fitting parameters is $N_\mathrm{para}=13+15\times(n+2)$. For simplicity, we assume KOI-977B's mass 
to be $M_\mathrm{B}=1.0M_\odot$. The stellar mass $M_\mathrm{B}$ is only relevant 
to Doppler boosting through Equation (\ref{eq:K}), 
and thus has little impact on the estimate for system parameters in the present case.

Due to the large number of fitting parameters, we found that the convergence 
of the fit was slow in case that we simply implement a burn-in with MCMC algorithm. 
In addition, the coefficients of polynomials for SC light-curve segments are sometimes
captured at a local minimum when we start the chain from an arbitrary set of initial guess 
for the coefficients. 
We thus employ the following procedure to minimize the $\chi^2$ statistics 
and estimate the uncertainties of the fitting parameters. First, we compute the initial guess
for the coefficients of polynomials ($15\times(n+1)$ parameters) by analytically solving the normal 
equations using the {\it out-of-transit} data for each of the SC light-curve segments. 
Then, allowing all the parameters to be free, we next minimize $\chi^2$ by 
Powell's conjugate direction method \citep[e.g.,][]{2002nrc..book.....P}. 
Starting with those optimized parameters, we implement MCMC algorithm to estimate the posterior 
distribution and look for possible correlations between free parameters. 
The best-fit value and its uncertainty are estimated from the median, and 15.87 and 84.13 percentiles
of the marginalized posterior distribution for each fitting parameter.

\begin{figure}[t]
\begin{center}
\includegraphics[width=9cm,clip]{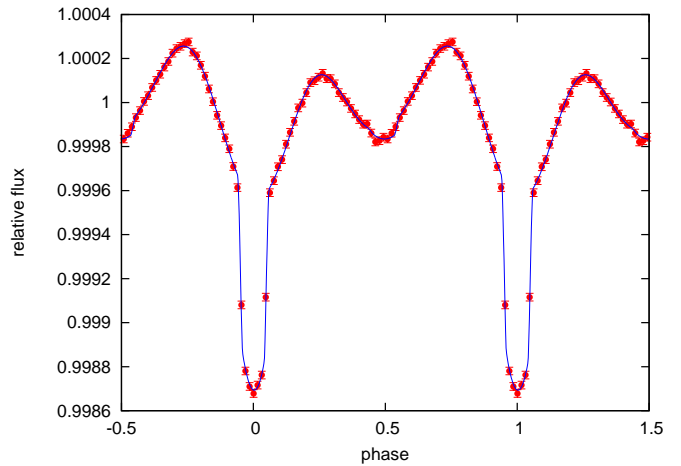} 
\caption{Phase-folded LC light-curve (red points with errorbars) along with the best-fit 
model (blue line) for $D=37$. 
}\label{fig:phasebest}
\end{center}
\end{figure}

\begin{figure}[t]
\begin{center}
\includegraphics[width=9cm,clip]{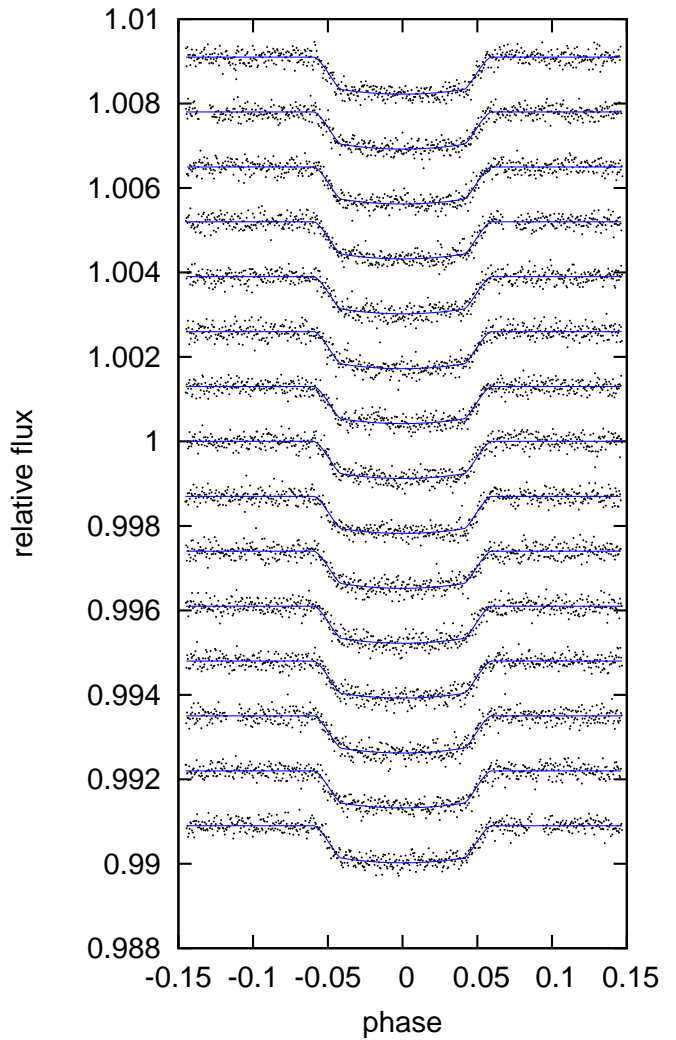} 
\caption{SC data around the transits in Q8 (black dots) along with the 
best-fit transit model for $D=37$. The baseline function ($5-$th order polynomial)
is subtracted from the original light-curve. 
}\label{fig:allsc}
\end{center}
\end{figure}

%%%%%%%%%%%%%%%%%%%%%%%%%%%%%%%%%%%%%%%%%%%%%%%%%%%%%%%%%%%%%%%%%%%%%%
\begin{table*}[tb]
\begin{center}
\caption{System Parameters for KOI-977B and KOI-977.01}\label{tab:bestfit}
%\begin{tabular}{lcl}
\begin{tabular}{lrrr}
\hline
Parameter& value ($D=37$) & value ($D=30$) & value ($D=20$)\\\hline\hline
{\it (Fitting Parameters)} &&& \\
$a/R_\mathrm{B}$ & $ 3.203_{-0.015}^{+0.014}$ & $3.169_{-0.023}^{+ 0.019}$ & $ 2.921 _{- 0.027 }^{+ 0.038 }$ \\ 
$b$ & $<5.3 \times 10^{-6}$ & $0.076_{-0.047}^{+ 0.068}$ & $ 0.425 _{- 0.030 }^{+ 0.024 }$ \\ 
$u_1+u_2$ & $ 0.210_{-0.088}^{+0.073}$ & $0.394_{-0.087}^{+ 0.077}$ & $ 0.357 _{- 0.069 }^{+ 0.066 }$ \\ 
$u_1-u_2$ & $ 0.520_{-0.195}^{+0.213}$ & $0.166_{-0.201}^{+ 0.205}$ & $ 0.286 _{- 0.216 }^{+ 0.239 }$ \\ 
$q$ & $ 0.277_{-0.011}^{+0.011}$ & $0.2096_{-0.0072}^{+ 0.0083}$ & $ 0.1147 _{- 0.0042 }^{+ 0.0044 }$ \\ 
$F_0$ & $ -0.00104_{-0.00031}^{+0.00031}$ & $-0.00076_{-0.00035}^{+ 0.00030}$ & $ -0.00038 _{- 0.00019 }^{+ 0.00021 }$ \\ 
$F_1$ & $ 0.00433_{-0.00013}^{+0.00012}$ & $0.00348_{-0.00011}^{+ 0.00011}$ & $ 0.00222 _{- 0.00007 }^{+ 0.00008 }$ \\ 
$\Delta\phi$ & $ 0.1500_{-0.0060}^{+0.0055}$ & $0.1500_{-0.0070}^{+ 0.0069}$ & $ 0.1529 _{- 0.0067 }^{+ 0.0074 }$ \\ 
$R_\mathrm{01}/R_{\rm B}$ & $ 0.17394_{-0.00079}^{+0.00076}$ & $0.15526_{-0.00083}^{+ 0.00090}$ & $ 0.12926 _{- 0.00073 }^{+ 0.00070 }$ \\ 
$e\cos\omega$ & $ -0.0110_{-0.0031}^{+0.0031}$ & $-0.0111_{-0.0034}^{+ 0.0034}$ & $ -0.0109 _{- 0.0033 }^{+ 0.0036 }$ \\ 
$e\sin\omega$ & $ 0.0071_{-0.0039}^{+0.0042}$ & $0.0035_{-0.0051}^{+ 0.0049}$ & $ 0.0035 _{- 0.0048 }^{+ 0.0048 }$ \\ 
$\Delta T_c^\mathrm{(LC)}$ (day) & $ 0.00088_{-0.00022}^{+0.00023}$ & $0.00093_{-0.00027}^{+ 0.00025}$ & $ 0.00090 _{- 0.00027 }^{+ 0.00026 }$ \\ 
\hline
{\it (Derived Parameters)} &&& \\
$I_o$ (deg) & $>89.9999$ & $88.63_{-1.26}^{+0.86}$ & $81.60_{-0.54}^{+0.70}$\\
$e$ & $0.0134_{-0.0033}^{+0.0036}$ & $0.0125_{-0.0034}^{+0.0036}$ & $0.0122_{-0.0035}^{+0.0036}$ \\
$\omega$ (deg) & $146_{-14}^{+18}$ & $162_{-21}^{+26}$ & $162_{-21}^{+25}$ \\
$\chi^2_\mathrm{best}$ & 9428.57 & 9410.60 & 9411.73 \\
\hline
\end{tabular}
\end{center}
\end{table*}
%%%%%%%%%%%%%%%%%%%%%%%%%%%%%%%%%%%%%%%%%%%%%%%%%%%%%%%%%%%%%%%%%%%%%%
We try several values for the order of polynomials $n$. 
In order to avoid overfitting the baseline, we compute Bayesian Information 
Criteria (BIC) for each $n$ by the following equation:
%%%%%%%%%%%%%%%%%
\begin{eqnarray}
\mathrm{BIC}=\chi_\mathrm{best}^2+N_\mathrm{para}\ln(N_\mathrm{data}),
\end{eqnarray}
%%%%%%%%%%%%%%%%%
where $N_\mathrm{data}$ is the number of data points. 
After fitting the LC and SC light-curves, we obtain $\mathrm{BIC}$ of
10537.40, 10499.04, and 10578.21 for $n=4$, 5, and 6, respectively,
with $N_\mathrm{data}$ being 8706. 
Therefore, $n=5$ seems to be the best case
for the baseline model and we adopt the best-fit parameters for $n=5$
as the final result summarized in the second column of Table \ref{tab:bestfit}. 
Figures \ref{fig:phasebest} and \ref{fig:allsc} plot the phase-folded LC data,
and all SC data around the fifteen transits, respectively, for $D=37$ and $n=5$. 
The best-fit models are shown in blue solid lines. In Figure \ref{fig:allsc}, 
each transit of KOI-977.01 has a vertical offset of 0.0013 for clarity. 
Note that the $\chi^2$ statistics in Equations (\ref{eq:chisq}) - (\ref{eq:chisqsc})
suggest that the fit of the transit-curve is almost dominated
by the SC datasets, and LC data have a very minor contribution in determining the
transit parameters. On the other hand, LC data are responsible for 
the other parameters like the mass ratio $q$ and emission/reflection of the secondary. 
Both fitting results in Figures \ref{fig:phasebest} and \ref{fig:allsc}
are compatible with one another (e.g., the transit depth), which justifies
our treatment.

%%%%%%%%%%%%%%%%%%%
%%%%%%%%%%%%%%%%%%%
%%%%%%%%%%%%%%%%%%%
%%%%%%%%%%%%%%%%%%%
Assuming that KOI-977B is a Sun-like star (with $M_\mathrm{B}\sim 1M_\odot$ and 
$R_\mathrm{B}\sim 1R_\odot$), 
the mass and radius ratios in Table \ref{tab:bestfit} indicate that KOI-977.01 is an M dwarf
rather than a planet or brown dwarf. The depth of secondary eclipse approximately gives 
the flux ratio between KOI-977.01 and KOI-977B, which is roughly estimated as $F_0+F_1\simeq 0.0033$
from Equation (\ref{eq:planetlight}). This is consistent with a typical flux ratio between 
a solar-type star and an M dwarf in the {\it Kepler} band. The negative value of $F_0$ 
is ascribed to the large value of $F_1$ (to keep $F_0+F_1$ a constant), which is likely 
caused by the large difference between the two flux peaks at $\phi\sim 0.25$ and $\phi\sim 0.75$. 
As we noted in Section \ref{s:lcmodel}, KOI-977.01 should be highly heated by KOI-977B
due to its proximity, and have an inhomogeneous intensity distribution 
on its surface, even if it is self-luminous. Therefore, whether tidally locked or not, KOI-977.01 
can have a flux variation as a function of $\phi$. Our model for this variation is
a simple sinusoidal model with small corrections for the orbital eccentricity and 
phase-shift, which may be responsible for the unphysical value for $F_0$.  
A precise modeling of the scattered light and global map of surface intensity 
of close-in binaries would settle this issue. 

Our best-fit model for $D=37$ indicates a vanishingly small impact parameter $b$. 
This is because the SC transit data imply a relatively short duration of
ingress and egress, and transit light-curves are more like U-shaped rather than V-shaped
which is frequently seen for eclipsing binaries. Figure \ref{fig:allsc} suggests that the
observed duration of ingress or egress is only $\sim 1/6$ of the duration of full transit, 
which is nearly equal to the radius ratio $R_\mathrm{01}/R_\mathrm{B}$ constrained by
the transit depth. This fact suggests that the dilution factor of $D=37$ is almost
the upper limit to account the observed transit light-curve; $D$ exceeding $\sim 40$
would result in a larger $R_\mathrm{01}/R_\mathrm{B}$ from the transit depth, 
and lead to a more V-shaped transit light-curve. 

Since the dilution factor $D$ is highly uncertain for the {\it Kepler} band, we also
try different values and estimate best-fit parameters for each
case. Our visual inspection of KOI-977's spectrum (without the I$_2$ cell) suggests
that the spectroscopic contamination should not be so considerable ($D\gtrsim 20$), 
so that we decide to try the cases of $D=30$ and 20. 
The fitting results for $D=30$ and 20 are also summarized in the third and fourth
columns in Table \ref{tab:bestfit}. As expected,
both mass and radius ratios for smaller $D$ take smaller values, but still reside in 
the mass regime of late-type stars. 
According to the best $\chi^2$ values in Table \ref{tab:bestfit}, along with the 
reasonable values for $b$, the models with smaller $D$ are favored than the case of $D=37$. 
A future deep direct imaging in different bands will allow a precise determination
of $D$.

%%%%%%%%%%%%%%%%%%%%%%%%%%%%%%%%%%%%%%%%%%%%%%%%%%%%%%%%%%%%%%%%%%%%%%
%\section{Global Analysis \label{s:global}}\label{s:global}
%Now that we have all the reduced data, we simultaneously fit the RV, binned LC light-curve, 
%and segmented SC light-curve. The $\chi^2$ statistics in the present case is expressed as
%%%%%%%%%%%%%%%%%
%\begin{eqnarray}
%\chi^2 =\chi^2_\mathrm{RV} + \chi^2_\mathrm{LC} + \chi^2_\mathrm{SC}.
%\end{eqnarray}
%%%%%%%%%%%%%%%%%
%Since KOI-977 is an evolved star and
%its light-curve also shows a signal of strong stellar pulsation (Figure \ref{fig:sc}), 
%RV variations due to pulsation is expected to be fairly large. Based on the empirical relation 
%for RV variations of evolved stars by \citet{2008A&A...480..215H}, we adopt the RV jitter for KOI-977 as 
%$\sigma_\mathrm{jitter}=100$ m s$^{-1}$, which is quadratically added to $\sigma_\mathrm{RV}$. 

%%%%%%%%%%%%%%%%%%%%%%%%%%%%%%%%%%%%%%%%%%%%%%%%%%%%%%%%%%%%%%%%%%%%%%
\section{Discussion and Summary \label{s:discussion}}\label{s:discussion}
In this paper, we have performed a global analysis for KOI-977 using 
spectroscopy, asteroseismology, and phase-curve modeling with 
{\it Kepler} light-curve data. Both spectroscopic and asteroseismic 
analyses indicate that KOI-977 is a giant with radius larger than 
$\sim20R_\odot$. Hence, the orbital period of KOI-977.01 implies
that it is not orbiting the most luminous giant (KOI-977A), but likely a companion 
to another star (KOI-977B) eclipsing each other. 
The light-curve folded with KOI-977.01's period shows a clear modulation
due to ellipsoidal variation. Assuming different values of the flux ratio between 
KOI-977A and 977B (the dilution factor $D$), we modeled the {\it Kepler} light-curves 
and extracted system parameters such as the mass and radius ratios. 
For $D\gtrsim20$, our fit to SC and folded LC light-curves
indicates that the mass and radius of KOI-977.01 are those of an M dwarf
assuming that KOI-977B is a Sun-like star. The orbital eccentricity of KOI-977.01
is estimated to be very small ($e<0.017$), which suggests a damping 
of $e$ by tidal interaction.

%%%%%%%%%%%%%%%%%%%
\begin{figure}[t]
\begin{center}
\includegraphics[width=9cm,clip]{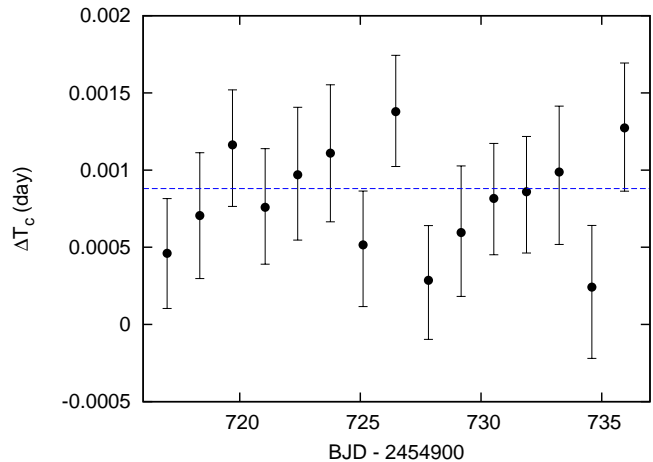} 
\caption{Deviation of the transit center from {\it Kepler}'s official ephemeris
for each transit of Q8, during which SC data were taken. No significant
TTV is seen in the data. 
}\label{fig:ttv}
\end{center}
\end{figure}
%%%%%%%%%%%%%%%%%%%
In the light-curve analysis in Section \ref{sec:fitcurve}, we allowed the fifteen
mid-transit times in the SC data to vary, and searched for a possible TTV signal. 
Figure \ref{fig:ttv} shows the resulting $\Delta T_c$ as a function of the predicted 
transit center $T_c$ from {\it Kepler}'s official ephemeris. The dashed horizontal line
indicates the best-fit $\Delta T_c^\mathrm{(LC)}$ for the folded LC light-curve. 
The SC and LC data are all consistent, and no significant TTV is visible, suggesting
that there is no nearby circumbinary companion around KOI-977B and KOI-977.01.

So far, we have not assumed that the eclipsing binary (KOI-977B and KOI-977.01)
is bound to the pulsating giant KOI-977A: KOI-977A could comprise a triple system, or
a foreground/background source aligned along the line-of-sight by chance. 
When we assume that KOI-977 is actually a triple system, and that
KOI-977B and KOI-977.01 are both normal stars
sharing the same age and metallicity as those for KOI-977A, 
we can put a further constraint on the stellar parameters of the eclipsing
binary through an isochrone model. Since an isochrone for an arbitrary age
(e.g., 3.25 Gyr) gives a relation between the stellar mass and radius, 
we can estimate the mass and radius for each star in the eclipsing binary 
from the observed mass and radius ratios. We define the following $\chi^2$
statistics as a function of stellar masses:
%%%%%%%%%%%%%%%%%
\begin{eqnarray}
\label{eq:chi_mass}
\chi^2(M_\mathrm{B}, M_\mathrm{01}) &=&
\frac{(q_\mathrm{obs}-M_\mathrm{01}/M_\mathrm{B})^2}{\sigma_q^2}\nonumber\\
&&+\frac{((R_\mathrm{01}/R_\mathrm{B})_\mathrm{obs}-R_\mathrm{01}/R_\mathrm{B})^2}{\sigma_{R_\mathrm{01}/R_\mathrm{B}}^2},
\end{eqnarray}
%%%%%%%%%%%%%%%%%
where $q_\mathrm{obs}$ and $(R_\mathrm{01}/R_\mathrm{B})_\mathrm{obs}$
are the observed mass and radius ratios in Table \ref{tab:bestfit}, and 
$\sigma_q$ and $\sigma_{R_\mathrm{01}/R_\mathrm{B}}$ are their errors, respectively. 
The radius $R_\mathrm{B}$ ($R_\mathrm{01}$) is a function of 
$M_\mathrm{B}$  ($M_\mathrm{01}$) via isochrone.

\begin{figure}[t]
\begin{center}
\includegraphics[width=9cm,clip]{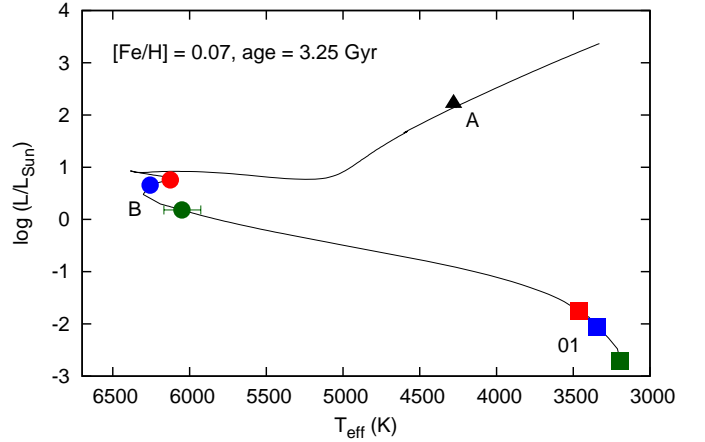} 
\caption{Dartmouth isochrone on the HR diagram along with the plots for the giant KOI-977A 
and eclipsing binary (KOI-977B and 977.01). Different colors correspond to different
assumed values of the dilution factor: $D=37,~30$, and 20 correspond to red, blue, and green,
respectively. 
}\label{fig:HRdiagram}
\end{center}
\end{figure}

%%%%%%%%%%%%%%%%%%%%%%%%%%%%%%%%%%%%%%%%%%%%%%%%%%%%%%%%%%%%%%%%%%%%%%
\begin{table}[tb]
\begin{center}
\caption{Stellar Parameters of KOI-977B and KOI-977.01 for the case that
their ages and metallicities are the same as KOI-977A}\label{tab:binary}
%\begin{tabular}{lcl}
\begin{tabular}{lrrr}
\hline
Parameter& value ($D=37$) & value ($D=30$) & value ($D=20$)\\\hline\hline
{\it (KOI-977B)} &&& \\
$M_\mathrm{B}~(M_\odot)$ & $1.402_{-0.014}^{+0.011}$ & $1.342_{-0.019}^{+0.017}$ & $1.095_{-0.048}^{+0.041}$ \\
$R_\mathrm{B}~(R_\odot)$ & $2.124_{-0.081}^{+0.071}$ & $1.811_{-0.068}^{+0.074}$ & $1.122_{-0.090}^{+0.086}$ \\
$T_\mathrm{eff}$ (K)& $6124_{-20}^{+38}$ & $6257_{-24}^{+5}$ & $6049_{-136}^{+122}$ \\
$\log g$ (dex) & $3.929_{-0.026}^{+0.029}$ & $4.049_{-0.029}^{+0.027}$ & $4.376_{-0.048}^{+0.054}$ \\
$\log(L/L_\odot)$ & $0.756_{-0.022}^{+0.024}$ & $0.656_{-0.032}^{+0.028}$ & $0.182_{-0.113}^{+0.099}$ \\
$M_{\rm Kp}$ (mag) & $2.67_{-0.06}^{+0.05}$ & $2.92_{-0.07}^{+0.08}$ & $4.10_{-0.25}^{+0.28}$ \\
\hline
{\it (KOI-977.01)} &&& \\
$M_\mathrm{01}~(M_\odot)$ & $0.386_{-0.018}^{+0.015}$ & $0.281_{-0.013}^{+0.014}$ & $0.125_{-0.010}^{+0.010}$ \\
$R_\mathrm{01}~(R_\odot)$ & $0.369_{-0.014}^{+0.012}$ & $0.281_{-0.010}^{+0.011}$ & $0.145_{-0.012}^{+0.011}$ \\
$T_\mathrm{eff}$ (K)& $3461_{-24}^{+21}$ & $3343_{-13}^{+14}$ & $<3198$ \\
$\log g$ (dex)& $4.883_{-0.012}^{+0.014}$ & $4.986_{-0.013}^{+0.011}$ & $5.211_{-0.031}^{+0.036}$ \\
$\log(L/L_\odot)$ & $-1.753_{-0.046}^{+0.039}$ & $-2.050_{-0.040}^{+0.042}$ & $-2.703_{-0.072}^{+0.064}$ \\
$M_{\rm Kp}$ (mag) & $9.86_{-0.11}^{+0.13}$ & $10.68_{-0.12}^{+0.11}$ & $12.45_{-0.16}^{+0.18}$ \\
\hline
$\chi^2(M_\mathrm{B}, M_\mathrm{01})$ & 0.033 & 0.018 & 0.006 \\
\hline
\end{tabular}
\end{center}
\end{table}
%%%%%%%%%%%%%%%%%%%%%%%%%%%%%%%%%%%%%%%%%%%%%%%%%%%%%%%%%%%%%%%%%%%%%%
%Since the Y$^2$ isochrone model \citep{2001ApJS..136..417Y}
%only covers the mass range greater than $0.4M_\odot$, 
We here adopt the Dartmouth isochrone model \citep{2008ApJS..178...89D}
since this model also gives the absolute magnitude in the {\it Kepler} band
($M_{\rm Kp}$) for an arbitrary stellar mass, so that we can directly compare 
the contrast between KOI-977A and 977B. 
Allowing $M_\mathrm{B}$ and $M_\mathrm{01}$ to vary, we minimized
Equation (\ref{eq:chi_mass}) and estimated the stellar parameters
of KOI-977B and KOI-977.01 for each case of $D$. 
For simplicity, we fix the metallicity and age at $[\mathrm{Fe/H}]=0.07$ and 3.25 Gyrs,
respectively, and use the corresponding single isochrone when we convert the mass 
into radius. Table \ref{tab:binary} summarizes the result of our fit. For all cases of
$D$, our fit indicates that KOI-977B and KOI-977.01 are late-F and
early-M dwarfs, respectively. %, which justifies the light-curve analysis in Section \ref{s:model}. 
Figure \ref{fig:HRdiagram} plots the Dartmouth isochrone (black line) on the HR diagram, 
for $[\mathrm{Fe/H}]=0.07$ and $\mathrm{age}=3.25$ Gyrs. The locations of KOI-977A, 977B, 
and 977.01 are plotted by the triangle, circles, and squares, respectively. 
Different colors for KOI-977B and 977.01 mean different $D$, with red, blue, 
and green corresponding to $D=37$, 30, and 20, respectively.

%%%%%%%%%%%%%%%%%%%
%%%%%%%%%%%%%%%%%%%
The {\it Kepler} absolute magnitude for the best-fit model
of KOI-977A is $M_{\rm Kp}\approx -0.55$, while those for KOI-977B and 977.01 are 
listed in Table \ref{tab:binary}. 
%mass estimates listed in Table \ref{tab:binary} correspond to $m_\mathrm{Kp}=2.68$ 
%for $M_B=1.402M_\odot$, $m_\mathrm{Kp}=2.92$ for $M_B=1.342M_\odot$, 
%$m_\mathrm{Kp}=4.11$ for $M_B=1.095M_\odot$, respectively. 
The differences in $M_{\rm Kp}$ for KOI-977A
and 977B are translated into the dilution factors of $D_\mathrm{isochrone}\approx 19,~24$, and 71
for the assumed $D$ of 37, 30, and 20, respectively. Comparing these values, we speculate that the true 
dilution factor $D$ should be 
located between 20 and 30 (closer to 30). It is possible to fit the light-curve for various $D$
and iteratively estimate the stellar parameters of KOI-977B and 977.01 through the isochrone, 
or more directly to perform a more global analysis with $D$ being free. 
However, the calculation above is based on a rather strong assumption that KOI-977A and the
eclipsing binary are bound to each other composing a triple system, so that we do not try such an
analysis here.  The confirmation of the companion and its common proper motion 
by future direct imaging is important in order to apply the above discussion. 
We will also be able to discuss the physical association between KOI-977A and 977B 
by measuring the absolute RV of KOI-977B with a near infrared spectroscopy using 
adaptive optics, and comparing the value with KOI-977A's absolute RV.

One remaining issue on the above scenario is that 
the phase-folded light-curve shows an asymmetric flux maximum that cannot
be explained by Doppler boosting. 
Our empirical treatment adopting a phase-shift in the emission/reflection of the secondary
resulted in a westward phase-shift of $\Delta \phi\sim 0.15$, which
is acceptable for a close-in giant planet (or brown dwarf) with an inhomogeneous cloud 
coverage on the surface \citep{2014arXiv1407.2245E}. However, 
this is only the case when the surface is relatively cool ($T_\mathrm{eq}\lesssim 2500$ K)
so that the surface cloud could be generated. Our analysis for the case that KOI-977 is a triple 
system indicates that KOI-977.01 is an early M star with $T_\mathrm{eff}\gtrsim 3000$ K, 
which is too hot to form a surface cloud. 
In addition, the magnitude of phase-shift ($\sim 0.15$) is too large for a stellar secondary. 
%This contradiction may suggest
%that a different process to produce a phase-shift in the intensity distribution is
%at work on KOI-977.01, or simply that the eclipsing binary is not bound to KOI-977A
%and KOI-977.01 is a less massive object (e.g., a brown dwarf). 
%Further observations are required to settle this issue. 

%%%%%%%%%%%%%%%%%%%
\begin{figure}[t]
\begin{center}
\includegraphics[width=9cm,clip]{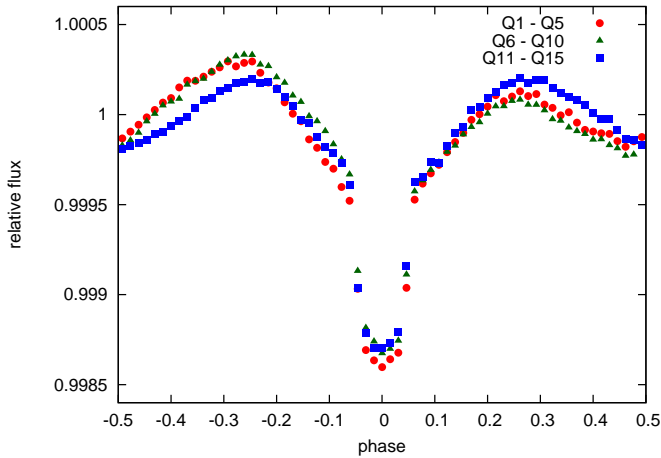} 
\caption{Variation in the phase-folded light-curve for five consecutive quarters. 
%The red circles, green triangles, blue squares represent the folded flux 
}\label{fig:lcQdiff}
\end{center}
\end{figure}
%%%%%%%%%%%%%%%%%%%
In order to gain a more insight into the asymmetric observed flux, 
we phase-fold the observed light-curves with only five consecutive quarters
instead of using the whole data. Figure \ref{fig:lcQdiff} shows the comparison 
among the folded light-curves for different epochs. 
The red circles, green triangles, and blue squares respectively represent the folded (and binned) 
fluxes for Q1--Q5, Q6--Q10, and Q11--Q15. This figure suggests that 
even if we fold the light-curve of a relatively long span ($>400$ days),
the folded flux can vary significantly; while folded light-curves 
for Q1--Q5 and Q6--Q10 exhibit the same feature (higher flux values at $\phi\sim -0.25$)
as in Figure \ref{fig:phasebest}, the one for Q11-Q15 looks very
different and the flux peaks at $\phi\sim -0.25$ and $\sim 0.25$ have approximately 
the same height. This variation of folded fluxes suggests that the impact of stellar 
pulsation may not be completely canceled out by phase-folding with a
timescale of a few years. This seems likely especially since the amplitude of KOI-977A's 
pulsation is $\sim 5\times 10^{-3}$ (see Figure \ref{fig:sc}), which is ten times 
greater than the amplitude of KOI-977B's ellipsoidal variation. 
Therefore, we suspect that the higher peak at $\phi\sim 0.75$ 
in Figure \ref{fig:phasebest} is not astrophysical, arising from an 
imperfect removal of stellar pulsation. 

If the higher peak at $\phi\sim -0.25$ is indeed an artifact (i.e., unphysical), the second term 
in the right-hand side of Equation (\ref{eq:planetlight}) should be interpreted as
an empirical correction term (by a sinusoidal fit) for the residual of KOI-977A's pulsation. 
A longer-term monitoring would uncover the origin of the apparent 
flux asymmetry in the observed data. It should be emphasized that 
the systematic errors arising from this artifact (e.g., in $q$) is much smaller than 
the uncertainty by the imperfect knowledge on $D$ (Table \ref{tab:bestfit}), 
and our conclusion that the eclipsing binary is likely comprised of F and M dwarfs
is not affected by the observed apparent asymmetry in the light-curve.

Our analysis has clarified that KOI-977.01 is very likely a false positive, suggesting that
the pulsating giant KOI-977A is different from the eclipsing binary that produces
transit-like signals. However, our technique which combines the information from
spectroscopy, asteroseismology, and light-curve analysis (transit, ellipsoidal variation, etc) 
was initially developed for a planet confirmation and its characterization. 
In fitting ellipsoidal variations, parameters such as 
$a/R$, $I_o$, and $q$ often have a degeneracy, and different sets
of the parameters yield similar light-curves \citep[see e.g., Equation (2) in][]{2011AJ....142..195S}. 
This degeneracy could be resolved by combining the phase-curve modeling with 
RV measurements and/or transit fits (for transiting systems). 
Basic stellar parameters estimated by asteroseismology and/or spectroscopy
would further help to characterize the systems (e.g., ages and stellar obliquities). 
We hope to present somewhere else a new report of planet discoveries by the 
global analysis developed here.

%%%%%%%%%%%%%%%%%%%%%%%%%%%%%%%%%%%%%%%%%%%%%%%%%%%%%%%%%%%%%%%%%%%%%%
\acknowledgments 

This paper is based on data collected at Subaru Telescope, which is
operated by the National Astronomical Observatory of Japan.  
We acknowledge the support for our Subaru HDS observations by Akito
Tajitsu, a support scientist for the Subaru HDS.  
T.H.\ is grateful to Norio Narita and Kiyoe Kawauchi for helping the
observation at Subaru. We express special 
thanks to Masayuki Kuzuhara and Hajime Kawahara for fruitful discussions on this subject. 
The data analysis was in part carried out on common use data analysis computer system 
at the Astronomy Data Center, ADC, of the National Astronomical Observatory of Japan.  
T.H.\ and O.B.\ are supported by Japan Society for Promotion of Science (JSPS) 
Fellowship for Research (No. 25-3183 and No. 25-13316). 
K.M.\ is supported by JSPS Research
Fellowships for Young Scientists (No. 26-7182) and by the
Leading Graduate Course for Frontiers of Mathematical
Sciences and Physics.
%J.N.W.\ and R.S.O.\ gratefully acknowledge support from the NASA Origins program
%(NNX11AG85G) and Kepler Participating Scientist program (NNX12AC76G).
%N.N.\ acknowledges support by the NAOJ Fellowship, the NINS Program
%for Cross-Disciplinary Study, and Grant-in-Aid for Scientific Research (A) (No. 25247026) 
%from the Ministry of Education, Culture, Sports, 
%Science and Technology (MEXT) of Japan.
We acknowledge the very significant cultural role and reverence that the
summit of Mauna Kea has always had within the indigenous people in Hawai'i. 
We express special thanks to the anonymous referee for the 
helpful comments and suggestions on this manuscript.

%%%%%%%%%%%%%%%%%%%%%%%%%%%%%%%%%%%%%%%%%%%%%%%%%%%%%%%%%%%%%%%%%%%%%%
%\appendix

%%%%%%%%%%%%%%%%%%%%%%%%%%%%%%%%%%%%%%%%%%%%%%%%%%%%%%%%%%%%%%%%%%%%%%
\bibliographystyle{apj} %added later
%\bibliography{benomar,hirano2014_reference}
%\end{document}
%%%%%%%%%%%%%%%%%%%%%%%%%%%%%%%%%%%%%%%%%%%%%%%%%%%%%%%%%%%%%%%%%%%%%%

%%%%%%%%%%%%%%%%%%%%%%%%%%%%%%%%%%%%%%%%%%%%%%%%%%%%%%%%%%%%%%%%%%%%%%

\end{document}